\documentstyle[twocolumn,pre,aps,epsf]{revtex}

\def\d{{\rm d}}
\def\e{{\rm e}}
\def\vector#1{{\bf #1}}

\def\vk{{\vector k}}
\def\vq{{\vector q}}

\def\vH{{\vector H}}

\def\dps{\displaystyle}
\def\vF{{v_{\rm F}}}
\def\Tc{{T_{\rm c}}}
\def\qbar{{\bar q}}
\def\hightc{{high-$T_{\rm c}$ }}

\def\kB{{k_{\rm B}}}
\def\Tc{{T_{\rm c}}}

\def\TMTSFPF{{${\rm (TMTSF)_2PF_6}$}}
\def\TMTSFClO{{${\rm (TMTSF)_2ClO_4}$}}

\def\RSGCO{{${\rm RuSr_2GdCu_2O_8}$}}

\begin{document}
\draft

\twocolumn[\hsize\textwidth\columnwidth\hsize\csname 
@twocolumnfalse\endcsname

\title{Upper critical fields of quasi-low-dimensional superconductors \\
with coexisting singlet and triplet pairing interactions \\
in parallel magnetic fields 
}

\author{Hiroshi Shimahara} 

\def\runtitle{}

\def\runauthor{Hiroshi {\sc Shimahara}} 

\address{
Department of Quantum Matter Science, ADSM, Hiroshima University, 
Higashi-Hiroshima 739-8526, Japan
}

\date{Received ~~~ March 2000}

\maketitle

\begin{abstract}
Quasi-low-dimensional type II superconductors in parallel magnetic fields 
are studied when singlet pairing interactions and relatively weak triplet 
pairing interactions coexist. 
Singlet and triplet components of order parameter are mixed at high fields, 
and at the same time an inhomogeneous superconducting state called 
a Fulde-Ferrell-Larkin-Ovchinnikov state occurs. 
As a result, the triplet pairing interactions enhance the upper critical 
field of superconductivity remarkably even at temperatures far above the 
transition temperature of parallel spin pairing. 
It is found that the enhancement is very large even when the triplet 
pairing interactions are so weak that a high field phase of parallel 
spin pairing may not be observed in practice. 
A possible relvance of the result in organic superconductors and 
a hybrid-ruthenate-cuprate superconductor is discussed. 
\end{abstract}



]

\narrowtext

Anisotropic superconductivities have been studied extensively in organic, 
oxide, and heavy fermion superconductors. 
For example, a triplet pairing is confirmed by Knight shift measurements 
in heavy fermion superconductors ${\rm UPt_3}$, 
while NMR experiments suggest a singlet pairing with a line node gap 
in ${\rm UPd_2Al_3}$. 
On the other hand, in a quasi-one-dimensional organic superconductor 
\TMTSFClO, a line node gap is supported by NMR data 
by Takigawa {\it et al.}~\cite{Tak87,Has87}, 
while thermal conductivity measurements by Belin {\it et al.} 
suggest a full gap superconductivity~\cite{Bel97,Shi00c}. 
In \TMTSFPF, triplet pairing superconductivity is supported 
by recent Knight shift measurements~\cite{Lee00}.

Pairing symmetries can be different in \TMTSFClO \hspace{0.25ex} and 
\TMTSFPF, but from the similarity of crystal structures and electronic 
states in these compounds, probably the pairing interactions have 
the same origin. 
From the phase diagram in pressure and temperature plane, 
$d$-wave like singlet superconductivity due to pairing interactions induced 
by the antiferromagnetic fluctuations have been discussed~\cite{Eme86,Shi89}. 
However, it has been discussed recently that such pairing interactions 
contain both singlet and triplet channels as attractive interactions, 
and even at zero field inter-site Coulomb interactions might favor 
a triplet pairing~\cite{Shi89,Shi00e}. 
Pairing interactions induced by antiferromagnetic fluctuations have been 
discussed also in \hightc oxide superconductors for the proximity to the 
antiferromagnetic phase~\cite{Miy88}.

In this paper, we examine quasi-low-dimensional superconductors 
in which singlet and triplet pairing interactions coexist. 
In particular, we concentrate on systems in which the singlet pairing 
interactions dominate at zero field. 
We calculate critical fields of superconductivity in directions parallel 
to the highly conductive layers. 
Because of the parallel direction, we assume that our system is strongly 
Pauli limited and the orbital pair-breaking effect can be ignored 
as a first approximation.

Matsuo and the present author {\it et al.} studied this problem in a three 
dimensional system with a spherical symmetric Fermi surface in which 
$s$-wave pairing interactions and weaker $p$-wave pairing interactions 
coexist~\cite{Mat94}. 
We found a remarkable enhancement of the critical field due to 
a mixing of order parameters of $s$-wave and $p$-wave symmetries. 
The order parameter mixing occurs due to appearance of non-zero 
center-of-mass momentum of Cooper pairs stabilized by Zeeman energy. 
Such an inhomogeneous superconducting state is called 
a Fulde-Ferrell-Larkin-Ovchinnikov state (FFLO or LOFF state). 
It should be noted that the enhancement occurs far above 
a transition temperature of the pure triplet pairing superconductivity 
which is estimated in the absence of the singlet pairing interactions. 
We discussed this effect in connection with the phase diagram of 
a heavy fermion superconductor~\cite{Glo93}. 
The order parameter mixing in the FFLO state have been pointed out 
also by Psaltakis {\it et al.} and Schopohl {\it et al.} in an $s$-wave 
superconductor and in a $p$-wave superfluid 
$^3{\rm He}$~\cite{Psa83,SchUP}.

In this paper, we extend our previous study to the two-dimensional 
systems and to the anisotropic singlet pairing. 
We assume inter-layer interactions implicitely so that the BCS-like 
mean field approximation is justified, while they are weak enough to be 
neglected in resultant mean field equations. 
In this sense, our systems are quasi-two-dimensional. 
We find that even very weak triplet pairing interactions enhance the 
critical fields remarkably also in the present systems.

The pairing interactions are expanded as 
\def\eqVexpand{(1)}
$$
     V(\vk,\vk') = - \sum_{\alpha} g_{\alpha} 
                     \gamma_{\alpha}(\vk) \gamma_{\alpha}(\vk') , 
     \eqno\eqVexpand
     $$
where $\gamma_{\alpha}(\vk)$ are defined by 
$\gamma_{d_{x^2-y^2}}(\vk) = {{\hat k}_x}^2 - {{\hat k}_y}^2$, 
$\gamma_{p_x}(\vk) = {\hat k}_x$ and so on 
in cylindrically symmetric systems. 
We take units with $\hbar = 1$ and $\kB = 1$ in this paper. 
We consider two cases: (i) $g_s > g_p > 0$ and (ii) $g_d > g_p > 0$, 
where we have defined $g_p \equiv g_{p_x} = g_{p_y}$ and 
$g_d \equiv g_{d_{x^2 - y^2}} = g_{d_{xy}}$. 
In each case, the other coupling constants are assumed to be zero. 
The gap function is expanded as 
\def\eqDeltaexpand{(2)}
$$
     \Delta (\vk) = \sum_{\alpha} \Delta_{\alpha} \gamma_{\alpha}(\vk) . 
     \eqno\eqDeltaexpand
     $$

The gap equations in the vicinity of the second order transition are 
written as 
\def\eqgapeq{(3)}
$$
     \Delta_{\alpha} = g_{\alpha} \sum_{\beta} 
                       K_{\alpha \beta} \Delta_{\beta} 
     \eqno\eqgapeq
     $$
with 
\def\eqKdef{(4)}
$$
     K_{\alpha \beta} = \frac{1}{N} \sum_{\vk'} 
     \gamma_{\alpha}(\vk') 
     \gamma_{\beta}(\vk') 
     \frac{1}{2} \sum_{\sigma} 
     \frac{\tanh{\frac{\epsilon_{\vk'} + \zeta \sigma}{2T}}}
          {2 \epsilon_{\vk'}} , 
     \eqno\eqKdef
     $$
where we have defined $\zeta = h (\qbar \cos \theta + 1)$, 
the angle $\theta$ between $\vk'$ and $\vq$, $\qbar = \vF q/2h$ and 
$h = |\mu_e \vH|$ with the electron magnetic moment $\mu_e$, 
It is apparent from the above equations that the mixing of the order 
parameter components of odd and even parities occurs only 
when both conditions of $\vq \ne 0$ and $h \ne 0$ are satisfied, 
which is expected from a symmetry consideration in momentum 
and spin spaces.

In the weak coupling limit, eq.{\eqgapeq} are rewritten in the form 
\def\eqgapeqweakcoupling{(5)}
$$
     \Delta_{\alpha} \log \frac{T}{T_{{\rm c}\alpha}^{(0)}}
     = - \sum_{\beta} M_{\alpha \beta} \Delta_{\beta}
     \eqno\eqgapeqweakcoupling
     $$
with 
$T_{{\rm c}\alpha}^{(0)} = {2 \e^\gamma \omega_{\rm c}}/{\pi} \cdot 
e^{- 1/g_{\alpha} N_{\alpha}(0)}$ and 
\def\eqMdef{(6)}
$$
\renewcommand{\arraystretch}{2.5}
     \begin{array}{rcl}
     M_{\alpha \beta} & \equiv & \dps{ 
       \int \frac{\d \varphi}{2\pi}
         \frac{\rho_{\alpha \beta}(0,\varphi)}{N_{\alpha}(0)}
         \sinh^2 \frac{\beta \zeta}{2} \, \Phi(\varphi)
     } \\
     \Phi(\varphi) & \equiv & \dps{ 
     \int_0^{\infty} \d y \log y 
       {\Bigl [} 
           \frac{2 \sinh^2 y}
                {(\cosh^2 y + \sinh^2 \frac{\beta\zeta}{2})^2}
     }\\ && \dps{ 
         - \frac{1}
                {\cosh^2 y \cdot ( \cosh^2 y + \sinh^2 \frac{\beta\zeta}{2} )}
       {\Bigr ]}
     } 
     \end{array}
     \eqno\eqMdef
     $$
where 
\def\eqrhodef{(7)}
$$
\renewcommand{\arraystretch}{2.0}
     \begin{array}{rcl}
     \rho_{\alpha \beta}(0,\varphi) & = & \dps{ 
     \gamma_{\alpha}(\varphi) \gamma_{\beta}(\varphi) \rho(0,\varphi) 
     } \\
     N_{\alpha}(0) & = & \dps{ 
     \int \frac{\d \varphi}{2\pi} \rho_{\alpha \alpha}(0,\varphi) . 
     } 
     \end{array}
     \eqno\eqrhodef
     $$
Here, $\varphi$ is the angle between $\vk'$ and $k_x$-axis, 
and $\rho(0,\varphi)$ is the angle dependent density of states 
at the Fermi energy. 
Equations {\eqgapeqweakcoupling} give the upper critical fields 
$h=h(T,\vq)$ for a given $\vq$. 
The final result of the critial fields is obtained by maximizing 
$h(T,\vq)$ with respect to $\vq$.

\vspace{\baselineskip}

\begin{figure}[htb]
\begin{center}
\leavevmode \epsfxsize=7cm  \epsfbox{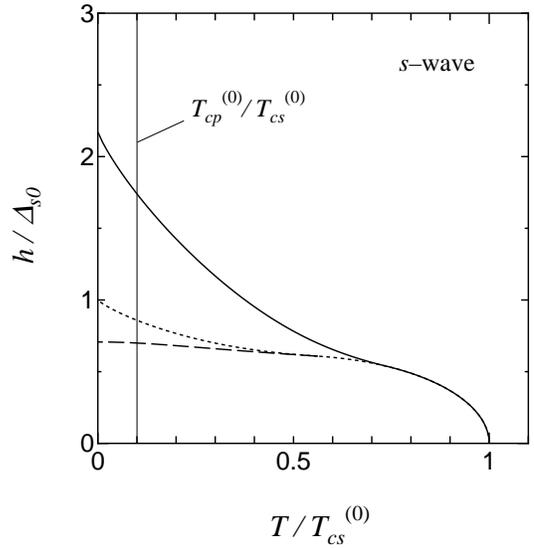}
\end{center}
\caption{
The upper critical fields of superconductivity in the case (i) 
in the presence (solid line) and the absence (dotted line) of the $p$-wave 
pairing interactions. 
$T_{{\rm c}p}^{(0)}/T_{{\rm c}s}^{(0)} = 0.1$ is assumed. 
The vertical thin solid line is the transition temperature of the pure 
$p$-wave superconductivity of parallel spin pairing. 
The broken line shows the Pauli paramagnetic limit when the FFLO state 
is ignored. 
Here, we have defined $\Delta_{s0} \equiv 2\omega_{\rm c} \e^{-1/g_sN_s(0)}$. 
}
\label{fig:sp}
\end{figure}

For the case of $s$-$p$-mixing (i), we can choose the vector $\vq$ 
in an arbitrary direction from the symmetry of the system. 
Thus, let us assume the direction of $\vq$ in $k_x$-axis. 
Then, the $p$-wave component with $\Delta_{p_y} \sim {\hat k}_y$ 
is not mixed with the $s$-wave component, 
while $\Delta_{p_x} \sim {\hat k}_x$ can be mixid. 
Thus, we only need to calculate the smallest 
eigen value $\lambda$ of the $2 \times 2$ matrix 
\def\eqspmixingMatrix{(8)}
$$
     {\left (
     \begin{array}{cc}
     M_{ss} & M_{sp} \\
     M_{ps} & M_{pp} + G_{p} 
     \end{array}
     \right )}
     \eqno\eqspmixingMatrix
     $$
where we have defined 
\def\eqGpdef{(9)}
$$
     G_{p} \equiv \log \frac{T_{{\rm c}s}^{(0)}}{T_{{\rm c}p}^{(0)}} 
     = \frac{1}{g_{p}N_{p}(0)} - \frac{1}{g_{s}N_{s}(0)} . 
     \eqno\eqGpdef
     $$
The transition temperature is given by 
$\Tc = T_{{\rm c}s}^{(0)} \e^{-\lambda}$.

For the case of $d$-$p$-mixing (ii), we can use the symmetry to 
fix the orientation of $d$-wave order parameter. 
Thus, let us assume $\Delta_{d} \sim {\hat k}_x^2 - {\hat k}_y^2$. 
In the absence of $p$-wave pairing interactions, 
it is known that the optimum $\vq$ is in the direction of $k_x$-axis 
(or equivalently $k_y$-axis) at low temperatures, 
while it is in the direction of the line of $k_y = \pm k_x$ 
at high temperatures~\cite{Mak96,Shi97}. 
In the presence of the $p$-wave pairing interactions, the direction of 
$\vq$ is not known a priori. 
Thus, we should assume that the two $p$-wave components 
(${\hat k}_x$ and ${\hat k}_y$) of order parameter 
can be mixed with the $d$-wave components. 
Therefore, we calculate the smallest eigen value $\lambda$ 
of a $3 \times 3$ matrix with elements 
$M_{dd}, M_{dp_x}, M_{p_xp_x} + G_{p}, \cdots $
and so on, which gives the transition temperature by 
$\Tc = T_{{\rm c}d}^{(0)} \e^{-\lambda}$. 
Here, the parameter $G_p$ is defined by 
\def\eqGpdefdpmixing{(10)}
$$
     G_{p} \equiv \log \frac{T_{{\rm c}d}^{(0)}}{T_{{\rm c}p}^{(0)}} 
     = \frac{1}{g_{p}N_{p}(0)} - \frac{1}{g_{d}N_{d}(0)} 
     \eqno\eqGpdefdpmixing
     $$
similarly to eq.{\eqGpdef}.

\vspace{\baselineskip}

\begin{figure}[htb]
\begin{center}
\leavevmode \epsfxsize=7cm  \epsfbox{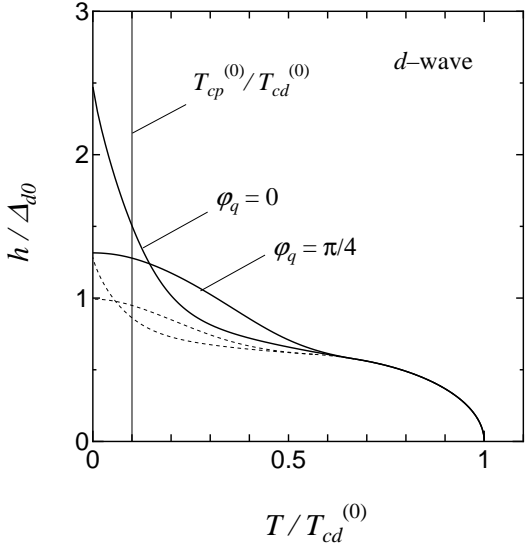}
\end{center}
\caption{
The upper critical fields in the case (ii) 
in the presence (solid line) and the absence (dotted line) 
of the $p$-wave pairing interactions. 
$T_{{\rm c}p}^{(0)}/T_{{\rm c}d}^{(0)} = 0.1$ is assumed. 
The vertical thin solid line is the transition temperature of the pure 
$p$-wave superconductivity of parallel spin pairing. 
$\varphi_q$ is the angle between $\vq$ and $k_x$-axis. 
At each temperature, the higher $h$ on the solid line is the final result 
of the critical field in the presence of the $p$-wave interactions. 
Here, we have defined $\Delta_{d0} \equiv 2\omega_{\rm c} \e^{-1/g_dN_d(0)}$. 
}
\label{fig:dpone}
\end{figure}

\begin{figure}[htb]
\begin{center}
\leavevmode \epsfxsize=7cm  \epsfbox{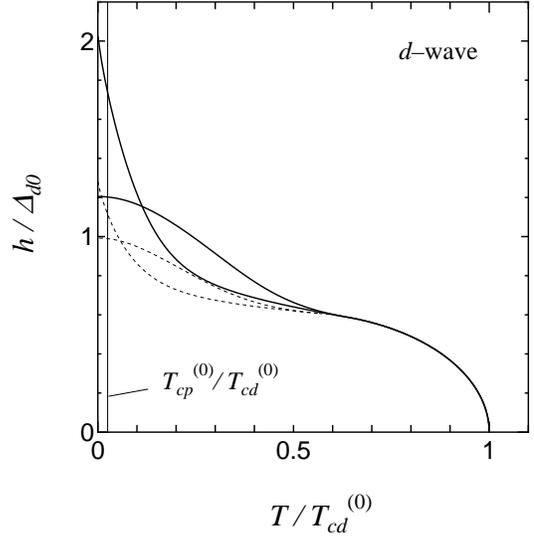}
\end{center}
\caption{
The upper critical fields for the case of $d$-$p$-mixing. 
$T_{{\rm c}p}^{(0)}/T_{{\rm c}d}^{(0)} = 0.025$ is assumed. 
The definitions of the lines are the same as in Fig.\ref{fig:dpone}. 
}
\label{fig:dptwo}
\end{figure}

\begin{figure}[htb]
\begin{center}
\leavevmode \epsfxsize=7cm  \epsfbox{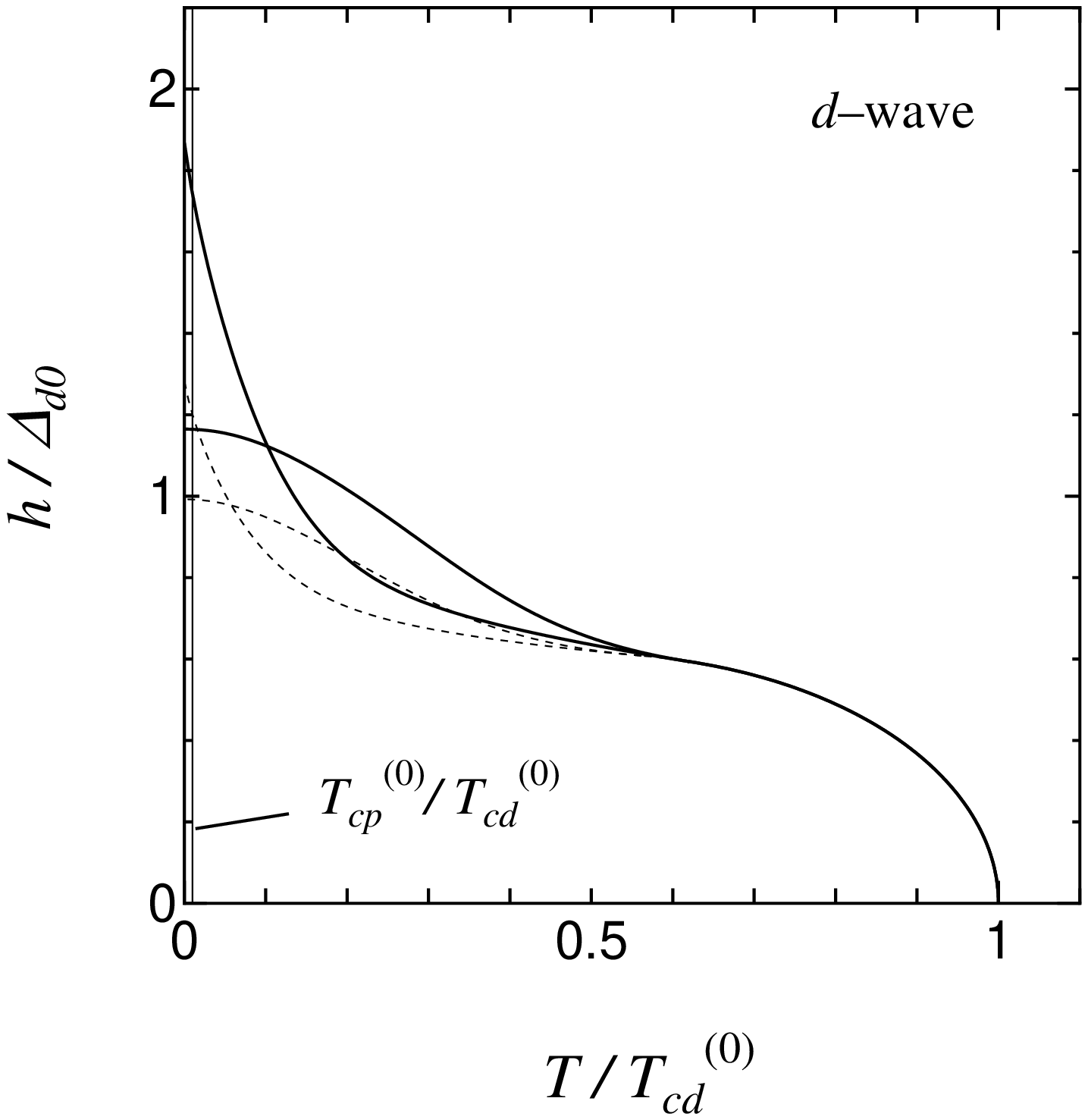}
\end{center}
\caption{
The upper critical fields for the case of $d$-$p$-mixing. 
$T_{{\rm c}p}^{(0)}/T_{{\rm c}d}^{(0)} = 0.01$ is assumed. 
The definitions of the lines are the same as in Fig.\ref{fig:dpone}. 
}
\label{fig:dpthree}
\end{figure}

Numerical results are drawn in figures \ref{fig:sp} - \ref{fig:dpthree}. 
Fig.\ref{fig:sp} shows the results for the $s$-$p$-mixing, while figures 
\ref{fig:dpone} - \ref{fig:dpthree} show those for the $d$-$p$-mixing. 
In the both cases, remarkable enhancements of the critical field are found 
at temperatures far above the transition temperature of the pure $p$-wave 
superconductivity of the parallel spin pairing. 
In particular, Fig.\ref{fig:dpthree} shows that 
the critical field is enhanced considerably even when 
$T_{{\rm c}p}^{(0)}/T_{{\rm c}d}^{(0)} = 0.01$.

For the $d$-$p$-mixing case, the optimum $\vq$ is in the direction of 
$k_x$-axis at low temperatures, while it is along the line of 
$k_y = \pm k_x$ at high temperatures, as shown in figures 
\ref{fig:dpone} - \ref{fig:dpthree}. 
In these figures, the results for $\varphi_q = 0$ and $\varphi_q = \pi/4$ 
are drawn by the solid lines. 
For each direction of $\vq$, the magnitude $q = |\vq|$ is optimized to 
obtain the critical field. 
It is confirmed by numerical calculations that $\vq$ with the other 
directions are not optimum.

\begin{figure}[htb]
\begin{center}
\leavevmode \epsfxsize=7cm  \epsfbox{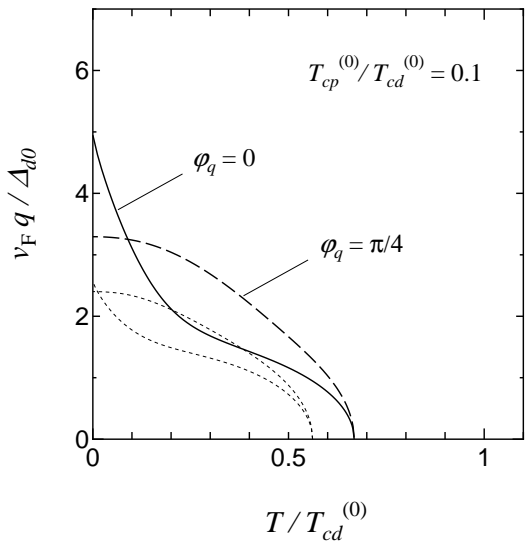}
\end{center}
\caption{
The magnitude of the optimum wave vector $\vq$ at the upper critical 
fields for $T_{{\rm c}p}^{(0)}/T_{{\rm c}d}^{(0)} = 0.1$ 
when $\varphi_q = 0$ (solid line) and $\varphi_q = \pi/4$ (broken line). 
The dotted lines are the result for pure $d$-wave pairing. 
}
\label{fig:qT}
\end{figure}

In Fig.\ref{fig:qT}, the optimum value of $q=|\vq|$ along the second 
order transition line is drawn. 
The temperature $T^{*}$ at which $q$ vanishes is the temperture of 
the tri-critical point of the normal phase, the BCS superconductivity 
($\vq = 0$) and the FFLO state ($\vq \ne 0$). 
It is found that the temperature $T^{*}$ increases due to the order 
parameter mixing in the presence of the $p$-wave interactions 
from the value $T^{*} \approx 0.561 \times T_{{\rm c}d}^{(0)}$ 
in the absence of the $p$-wave interactions. 
For example, $T^{*} \approx 0.668 \times T_{{\rm c}d}^{(0)}$ is 
estimated for $T_{{\rm c}p}^{(0)} / T_{{\rm c}d}^{(0)} = 0.1$.

\vspace{\baselineskip}

\vspace{\baselineskip}

\vspace{\baselineskip}

In conclusion, we examined upper critical field of the superconductivity 
when the singlet and triplet pairing interactions coexist. 
We extended our previous study in a spherical symmetric system~\cite{Mat94} 
to the quasi-two-dimensional systems in parallel magnetic fields and 
to the anisotropic singlet ($d$-wave) pairing. 
We obtained similar results to our previous results, and find that 
even very weak triplet intereactions enhance the critical fields 
remarkably, and the temperature of the tri-critical point of the normal, 
BCS, and the FFLO states is also enhanced. The present phase diagrams 
coincide with one predicted in our previous paper~\cite{Shi00e} 
except that the orbital pair-breaking effect is not taken into account 
in the present paper.

In a \hightc superconductor \RSGCO, for the coexistence of the 
superconductivity and the ferromagnetism (or canted 
ferromagnetism~\cite{Aki00}), 
the FFLO state has been discussed~\cite{Pic99,Shi00d}. 
Probably, $d$-wave pairing interactions are dominant in this system, 
but weak triplet pairing interactions must coexist because of 
the proximity to the magnetic phase. 
Therefore, the present mechanism which enhances the critical fields and 
$T^{*}$ may play a role in stabilizing a bulk superconductivity 
in this system~\cite{Shi00d}.

In the organic superconductors, the FFLO state has been discussed 
by many authors~\cite{Lee97,Sym99,Shi97a} 
to explain the high upper critical fields which exceed a conventional 
estimation of Pauli paramagnetic limit (Chandrasekar-Clogston limit). 
As we have discussed above, there is a possibility 
that the antiferromagnetic fluctuations contribute to 
the pairing interactions in these systems, 
and they should contain both singlet and triplet 
channels as attractive interactions~\cite{Shi00e}. 
Therefore, the present mechanism may contribute to stabilizing the 
superconducting phase at high fields. 
As shown in Fig.\ref{fig:dpthree}, the enhancement can be very large 
even when the triplet pairing interactions are so small that the pure 
triplet superconductivity of parallel spin pairing is not observed 
in practice.

This work was supported by a grant for CREST from JST.


\end{document}